\documentclass{osa-article}

\journal{oe}

\usepackage{amsmath,amssymb,graphicx}
\usepackage{color}
\usepackage{braket}

\articletype{Research Article}

\begin{document}

\title{Analysis of superresolution via 3D structured illumination intensity correlation microscopy}

\author{Anton Classen,\authormark{1,2,*} Joachim von Zanthier,\authormark{1,2} and Girish S. Agarwal\authormark{3}}

\address{\authormark{1}Institut f\"ur Optik, Information und Photonik, Universit\"at Erlangen-N\"urnberg, 91058 Erlangen, Germany\\
\authormark{2}Erlangen Graduate School in Advanced Optical Technologies (SAOT), Universit\"at Erlangen-N\"urnberg, 91052 Erlangen, Germany\\
\authormark{3}Institute for Quantum Science and Engineering and Department of Biological and Agricultural Engineering and Department of Physics and Astronomy, Texas A\&M University, College Station, TX 77843-4242, USA}

\email{\authormark{*}anton.classen@fau.de} %% email address is required

\begin{abstract} 
Intensity correlation microscopy (ICM), which is prominently known through antibunching microscopy or super-resolution optical fluctuation imaging (SOFI), provides superresolution through a correlation analysis of antibunching of independent quantum emitters or temporal fluctuations of blinking fluorophores. For correlation order $m$ the PSF in the signal is effectively taken to the $m$th power, and is thus directly shrunk by the factor $\sqrt{m}$. Combined with deconvolution a close to linear resolution improvement of factor $m$ can be obtained. Yet, analysis of high correlation orders is challenging, what limits the achievable resolutions. Here we propose to use three dimensional structured illumination along with $m$th-order correlation analysis to obtain an enhanced scaling of up to $m+m=2m$. Including the stokes shift or plasmonic sub-wavelength illumination enhancements beyond $2m$ can be achieved. Hence, resolutions far below the diffraction limit in full 3D imaging can potentially be achieved already with low correlation orders. Since ICM operates in the linear regime our approach may be particularly promising for enhancing the resolution in biological imaging at low illumination levels.
\end{abstract}

\section{Introduction}

Since it was first shown that the resolution limit, posed by diffraction, can be overcome \cite{Hell1994}, a variety of superresolution microscopy methods have been developed. Yet, each technique comes with certain requirements and limitations, thus justifying an ongoing pursuit of novel methods. One group of methods relies on stimulated ground or excited state depletion and a non-linear fluorophore response to deterministically engineer the effective excitation point spread function (PSF) \cite{Hell1994,Hell1995,Hell2005}. Other methods stochastically localize single photoswitchable molecules via centroid fitting of the PSF \cite{Moerner1997,Mason2006,Xiaowei2006,Betzig2006}.

Another branch of methods makes use of intensity correlations that are evaluated from an image series \cite{Dertinger2009,Schwartz2013,Genovese2014a}. For these intensity correlation microscopy (ICM) techniques, statistically blinking fluorophores \cite{Dertinger2009} or quantum emitters that exhibit anti-bunching \cite{Schwartz2013} can be used to enhance the resolution, both in widefield \cite{Schwartz2013} or confocal microscopy \cite{Genovese2014a}, by shrinking the effective PSF by the factor $\sqrt{m}$ (with correlation order $m$), and thus leading to a resolution improvement of up to factor $m$ when including deconvolution. Especially the first approach, known as superresolution optical fluctuation imaging (SOFI), is widely applied due to its combination of resolution improvement with low complexity of use \cite{Enderlein2016,Zhang2015,Dedecker2012}. Yet, in practice high correlation orders are not evaluated due to strong brightness skewing in the final image and long measurement times to obtain a reliable evaluation \cite{Enderlein2016}. Together with the moderate scaling of factor $m$, ICM currently does not provide resolutions far below the diffraction limit.

In parallel, structured illumination microscopy (SIM) was developed, where by the use of spatial frequency mixing the resolution is doubled within the linear wave optics regime \cite{Heintzmann1999,Gustafsson2000}. The non-linear derivative \textit{saturated} SIM leads to an in principle unlimited resolution, though at the cost of necessitating high intensities \cite{Heintzmann2002,Gustafsson2005}. Other derivatives combine SIM with the third-order process of CARS \cite{Rubinsztein2010,Lee2014}, or with surface plasmons \cite{Zubairy2014,Wei2014,Ponsetto2017} to access higher spatial frequency information. 3D SIM doubles both, the lateral and axial resolution \cite{Gustafsson2008}, while standing wave fluorescence microscopy techniques \cite{Bailey1993,Freimann1997} highly enhance the axial resolution via a dense axially structured illumination, but not the lateral one. Double-objective illumination and detection techniques \cite{Hell1994a,Gustafsson1999} with 3D-SIM attain the axial resolution of standing wave fluorescence microscopy and the lateral one of 2D SIM \cite{Shao2008}. Today, the SIM toolbox is considered to be one of the most powerful and versatile superresolution techniques, due to its combination of resolution improvement with good acquisition speed and flexibility of use \cite{Strohl2016}.

Recently, we showed that SIM and ICM based on antibunching can be combined to enhance the lateral resolution \cite{Classen2017a}. For correlation order $m$ the enhancement scales favorably as $m + m = 2m$ (when including deconvolution), which is a large improvement over the moderate factor $m$ scaling of antibunching microscopy itself. A similar result for 2D SIM combined with SOFI was later derived by Zhao \textit{et al.} \cite{Zhao2017}, resulting in structured illumination SOFI. While the two approaches make use of different physical processes (antibunching or statistical fluctuations) their final signals take the same form such that we identify them as structured illumination intensity correlation microscopy (SI-ICM). 

Here, we propose to use 3D structured illumination \cite{Gustafsson2008} in combination with ICM to equally enhance the axial resolution by the factor $2m$. This is a crucial step resulting in full 3D superresolution capability of SI-ICM. We present the theory and illustrate the basic flow chart of the technique. We point out that ICM and SIM operate within the linear regime and are established techniques in the field of superresolution microscopy. Thus, SI-ICM bears the potential for full 3D deep-subwavelength resolution at low illumination levels.

\section{Theory}
\label{sec:theo}

\begin{figure}[t]%
\centering
\includegraphics[width=1.0 \linewidth]{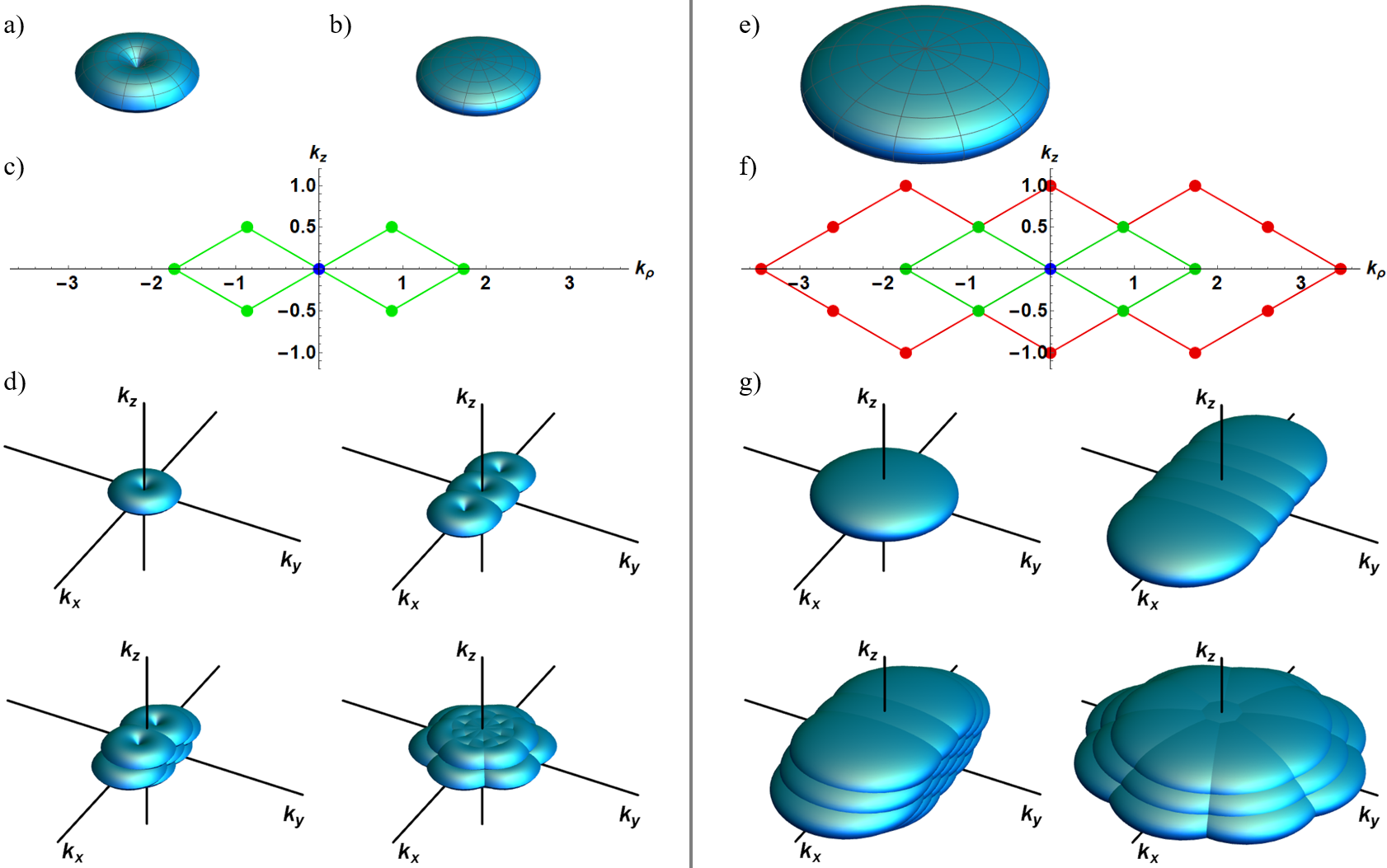}
\caption{Illustrations of total OTF supports of 3D-SIM (left column) and second-order 3D SI-ICM (right column). Image (a) depicts the OTF of a widefield microscope and (b) the  3D Gaussian $H(\mathbf{k})$ as approximation. (c) shows the Fourier transform of the structured illumination of Eq.\,(\ref{eq:8}) with the center positions $\mathbf{k}_j=(k_\rho, k_z)_j$ $(j=1,\ldots,7)$ given in the main text (see the central blue and outer green dots). Combining (a) and (c) yields the images in (d), where the OTF support of widefield microscopy, 2D SIM for one single orientation $\alpha$, 3D-SIM for one $\alpha$ and 3D-SIM for three orientations $\alpha = 0,\frac{\pi}{3},\frac{2\pi}{3}$ are shown. The final image of (d) provides a two-fold enlarged support along all axes. Image (e) shows the OTF $H_2(\mathbf{k})$ of second-order ICM, which is enlarged by the factor 2 along all axes. Image (f) depicts the Fourier transform of the squared structured illumination, where the outer (red) dots represent the contributions from the first higher harmonics. Again, combining images (e) and (f) yields the OTF supports displayed in g), i.e., of second-order ICM, second-order ICM with 2D-SIM for a single $\alpha$, second-order ICM with 3D SIM for a single $\alpha$ and second-order ICM with 3D-SIM for four orientations $\alpha = 0,\frac{\pi}{4},\frac{2\pi}{4},\frac{3\pi}{4}$. The total support for this case is already enhanced by the factor $4$ along all axes.}
\label{fig:3D-1}%
\end{figure}

Without loss of generality we assume $\mathbf{R} \equiv \mathbf{r}$ for the coordinates in the object and image plane, respectively, i.e. a magnification of one. Let $h(\mathbf{r})$ be the 3D PSF of a given widefield microscope, with $\mathbf{r} = (x,y,z)$. $H(\mathbf{k}) \equiv FT\{ h(\mathbf{r}) \}$ denotes the corresponding 3D optical transfer function (OTF) obtained by Fourier transform ($FT$) of $h(\mathbf{r})$, where $\mathbf{k} = (k_x,k_y,k_z)$ denotes the spatial frequency in reciprocal space. The lateral and axial widths of the PSF determine the resolution power a microscope provides to discern individual close-by emitters. The lateral width is usually smaller than the axial one. Moreover the axial resolution can not properly be defined in widefield microscopy, which is due to the missing z-cone in Fourier space [see Fig.\,\ref{fig:3D-1}(a)] \cite{BornWolf1999}. Optical sectioning capability in z-direction can however be retrieved by the measurement of z-stacks and deconvolution, using a pinhole as in confocal microscopy, or a variety of other means. Note that in ICM, the missing z-cone is intrinsically removed and thus true optical sectioning capability is provided already via the correlation analysis \cite{Dertinger2012,Schwartz2013}.

Here, as an approximation to the real widefield microscopy PSF, we consider a 3D Gaussian PSF of the form \cite{Dertinger2009}
\begin{equation}
h(\mathbf{r}) = \exp \left[-\frac{x^2 + y^2}{w_\rho^2} - \frac{z^2}{w_z^2}\right], 
\label{eq:1}
\end{equation}
where $w_\rho$ and $w_z$ denote the lateral and axial width, and $\rho = (x^2 + y^2)^{1/2}$. The OTF is then also a 3D Gaussian in reciprocal space [see Fig.\,\ref{fig:3D-1}(b)]. We point out that this is a useful assumption, since below the effective PSF will be the original PSF taken to the $m$th power $h_m(\mathbf{r}) \equiv h(\mathbf{r})^m$ and thus approaches a 3D Gaussian. The same is valid  for the OTF $ H_m(\mathbf{k}) \equiv FT\{ h_m(\mathbf{r}) \} = FT\{ h(\mathbf{r}) \} \ast \cdots \ast FT\{ h(\mathbf{r})\} = H(\mathbf{k}) \ast \cdots  \ast H(\mathbf{k})$ which is the $m$-fold convolution  of itself (denoted by `$\ast$') and equally approaches a 3D Gaussian in Fourier space. We consider the ratio $w_z / w_\rho = 3.0$ to mimic a typical widefield microscopy PSF, where the resolvable distances along $\rho$ and $z$ between two close-by emitters are given by \cite{Novotny2006}
\begin{equation}
\Delta \rho_\text{min} = 0.61 \frac{\lambda}{\mathcal{A}}  \qquad \Delta z_\text{min} = 2 \frac{n \lambda}{\mathcal{A}^2} \, .
\label{eq:2}
\end{equation}
In Eq.\,(\ref{eq:2}), \noindent $\mathcal{A}$ is the numerical aperture, $n$ the refractive index and $\lambda$ the wavelength of the emitted fluorescence light. To simplify the illustration, we set the stokes shift to zero, resulting in equal wavelengths for excitation and emission, i.e. $\lambda_\text{ex} = \lambda_\text{em}  \equiv \lambda$. Note though that $\lambda_\text{ex} \neq \lambda_\text{em}$ can easily be incorporated in the analysis.

The fluorophores are considered to be driven (far) below saturation resulting in a linear response to the (monochromatic) illumination intensity $I_{\text{str}}(\mathbf{r})$. In widefield microscopy and ICM a plane-wave illumination leads to the flat excitation intensity $I_{\text{str}}(\mathbf{r}) = I_0$. The model system to be imaged $n(\mathbf{r}) \propto \sum_{i=1}^N \delta(\mathbf{r}-\mathbf{r}_i)$ can be described by an ensemble of approximately point-like emitters at positions $\mathbf{r}_i$. Considering the convolution by the PSF the (time-averaged) image of this ensemble, taken in the image plane, reads  
\begin{equation}
\begin{aligned}
\braket{I(\mathbf{r},t)} \equiv I(\mathbf{r}) = h(\mathbf{r}) \ast n(\mathbf{r}) = I_0 \sum_{i=1}^N h(\mathbf{r}-\mathbf{r}_i) \, .
\end{aligned}
\label{eq:2b}
\end{equation}
$I_0$ denotes the average emitter intensity, which here is assumed to be equal for each emitter. Note that the intensity $I(\mathbf{r}) \equiv G^{(1)}(\mathbf{r})$ can also be recognized as Glauber's first-order equal-time intensity correlation function $G^{(1)}(\mathbf{r},t;\mathbf{r},t) = \langle \hat{E}^{(-)}(\mathbf{r},t) \hat{E}^{(+)}(\mathbf{r},t) \rangle$ assuming an ergodic system \cite{Glauber1963-2}. $\hat{E}^{(+)}$ and $\hat{E}^{(-)}$ are the positive and negative frequency parts of the electric field operator \cite{Classen2017a}.

Intensity correlation analysis can enhance the resolution, given that a certain process enables to discern and localize individual emitters within a sub-diffraction area or volume. For quantum emitters it is the intrinsic antibunching property that allows for an enhanced resolution \cite{Schwartz2013}, while for SOFI it is the independent and statistical blinking of fluorophores \cite{Dertinger2009}. The resulting final signals are however of the same form with the PSF being taken to the $m$th power. Since the detailed derivations of ICM can be found elsewhere \cite{Schwartz2013,Dertinger2009} we only provide a brief sketch here. First,  consider the squared (measured) intensity of Eq.\,(\ref{eq:2b}) 
\begin{equation}
[G^{(1)}(\mathbf{r})]^2 = I_0^2 \sum_{i, j=1}^N h(\mathbf{r}-\mathbf{r}_i) h(\mathbf{r}-\mathbf{r}_j) \equiv I_0^2  \sum_{i=1}^N h^2(\mathbf{r}-\mathbf{r}_i) + I_0^2 \sum_{i\neq j}^N h(\mathbf{r}-\mathbf{r}_i) h(\mathbf{r}-\mathbf{r}_j) \, .
\label{eq:g1Sq}
\end{equation}
Simply squaring the intensity does not provide superresolution, but in the first sum of Eq.\,(\ref{eq:g1Sq}) the squared PSF arises. It is only the second sum with the (detrimental) cross terms which prevents the entire signal to be superresolving. To isolate the terms with the squared PSF (without application of any \textit{a priori} knowledge) one can make use of the second-order intensity correlation function $G^{(2)}(\mathbf{r})   \equiv  G^{(2)}(\mathbf{r},\mathbf{r})  = \langle \hat{E}^{(-)}(\mathbf{r},t) \hat{E}^{(-)}(\mathbf{r},t) \hat{E}^{(+)}(\mathbf{r},t) \hat{E}^{(+)}(\mathbf{r},t)\rangle$ \cite{Glauber1963-2}. The correlation functions for the two different approaches read \cite{Schwartz2013,Dertinger2009,Classen2017a}
\begin{equation}
\begin{aligned}
\text{Antibunching:\quad }& G^{(2)}(\mathbf{r}) \propto I_0^2 \sum_{i\neq j}^N h(\mathbf{r}-\mathbf{r}_i) h(\mathbf{r}-\mathbf{r}_j) \\
\text{SOFI:\quad }& G^{(2)}(\mathbf{r}) \propto I_0^2 \sum_{i,j=1}^N h(\mathbf{r}-\mathbf{r}_i) h(\mathbf{r}-\mathbf{r}_j) + \sum_{i=1}^N h^2(\mathbf{r}-\mathbf{r}_i)\braket{\Delta I_i(t)^2} 
\end{aligned}
\label{eq:AB1}
\end{equation}
For antibunching, each individual source (e.g fluorophores, quantum dots, etc) can emit at most one photon per excitation cycle such that only cross terms survive. In SOFI, the blinking fluorophores $I_i(t) = I_0 + \Delta I_i(t)$, with zero-mean fluctuations $\Delta I_i(t)$, lead to the excess countrates $h^2(\mathbf{r}-\mathbf{r}_i)\braket{\Delta I_i(t)^2}$ compared to $[G^{(1)}(\mathbf{r})]^2$. A convenient subtraction of the terms in Eqs.\,(\ref{eq:g1Sq}) and (\ref{eq:AB1}) results in the sought-after (second-order) ICM signals \cite{Dertinger2009,Schwartz2013,Classen2017a}
\begin{equation}
\begin{aligned}
\text{Antibunching:\quad }  & \text{ICM}_2(\mathbf{r}) = \left[G^{(1)}(\mathbf{r})\right]^2 - G^{(2)}(\mathbf{r}) = I_0^2 \sum_{i=1}^N h^2(\mathbf{r}-\mathbf{r}_i) \, , \\ 
\text{SOFI:\quad }  & \text{ICM}_2(\mathbf{r}) = G^{(2)}(\mathbf{r}) - \left[G^{(1)}(\mathbf{r})\right]^2 = \overline{(\Delta I)^2} \sum_{i=1}^N h^2(\mathbf{r}-\mathbf{r}_i)  \, .
\end{aligned}
\label{eq:2a}
\end{equation}
where we considered $\overline{(\Delta I)^2} = \braket{\Delta I_i(t)^2}$ for each emitter. The squared PSF $h^2(\mathbf{r})$ directly leads to a resolution enhancement of factor $\sqrt{2}$. Though, the OTF $H_2(\mathbf{k}) = H(\mathbf{k}) \ast  H(\mathbf{k})$ is effectively twice as large. Hence, by rescaling the strongly suppressed Fourier amplitudes in the outer rims of the support the resolution can be doubled [depending on the signal-to-noise ratio (SNR)]. This process, known as deconvolution, is achieved trough application of a Wiener filter [see Eq.\,(\ref{eq:SIM1}) below]. Higher order $\text{ICM}_m(\mathbf{r})$ signals are evaluated in a similar manner as a combination of all correlation orders up to $m$ and read \cite{Dertinger2009,Schwartz2013,Classen2017a} 
\begin{equation}
\text{ICM}_m(\mathbf{r}) = \sum_{i=1}^N h_m(\mathbf{r}-\mathbf{r}_i) \, ,
\label{eq:3}
\end{equation}
where we set $I_0^m \equiv 1 $ and $\overline{(\Delta I)^m} \equiv 1$ for all orders $m$ to simplify the illustration. Note that the required combinations for SOFI are equivalent to so-called cumulants \cite{Dertinger2009}, and are different from the ones for antibunching microscopy which can be found in \cite{Schwartz2013}. Even though high correlation orders can in principle be evaluated, the moderate scaling with $m$ prevents resolutions far below the diffraction limit. In addition different molecular brightnesses and/or blinking ratios $\braket{\Delta I_i(t)}$ become more pronounced with rising order $m$, what skews the final image \cite{Enderlein2016}. While this problem is mitigated by use of balanced cumulants \cite{Geissbuehler2012}, in practice still only low correlation orders are utilized.

A different and independent approach to enhance the resolution of optical microscopy is SIM. Only recently it was realized that SIM and ICM can fruitfully be combined to enhance the lateral resolution \cite{Classen2017a,Zhao2017}. The goal of this manuscript is to show that the same holds true for the axial resolution and thus full 3D superresolution is possible via 3D SI-ICM. Towards this, we review  2D \cite{Heintzmann1999,Gustafsson2000} and especially 3D SIM \cite{Gustafsson2008} in detail, since this knowledge will be particularly helpful to understand the combination of SIM and ICM in Eq.\,(\ref{eq:6}) below. 

In 2D SIM two coherent plane waves are superposed at an angle which e.g. stem from the $\pm 1$ diffraction orders of a grating. The electric field distribution reads $E = e^{i  (k_x x + k_y y) + i k_z z} + e^{-i  (k_x x + k_y y) + i k_z z}$ resulting in the intensity pattern $I_{\text{str}}(\mathbf{r}) = |E|^2 = 1 + \cos[2  (k_x x + k_y y)]$. We denote the orientation of the pattern by $\alpha = \tan^{-1}(k_y/k_x)$. Moving the grating laterally (along $\alpha$) acts as opposite \textit{lateral} phase shifts $\pm \varphi_r $ on the two beams and thus leads to the pattern $I_{\text{str}}(\mathbf{r})= I_0 [  \frac{1}{2} + \frac{1}{2} \cos (\mathbf{k}_0 \mathbf{r} + 2 \varphi_r )]$, with $|\mathbf{k}_0| \equiv 2 (k_x^2 + k_y^2)^{1/2}$. Due to the illumination the individual emitter intensities are scaled by $I_{\text{str}}(\mathbf{r}_i)$, what results in measured images of the form \cite{Gustafsson2000}
\begin{equation}
%\begin{aligned}
\text{SIM}(\mathbf{r})  = h(\mathbf{r}) \ast \left[ n(\mathbf{r})  \times I_{\text{str}}(\mathbf{r},\alpha,\varphi_r) \right] = \sum_{i=1}^N  h(\mathbf{r}-\mathbf{r}_i)  \times I_{\text{str}}(\mathbf{r}_i,\alpha,\varphi_r) 
\, .
%\end{aligned}
\label{eq:5}
\end{equation}
The multiplication $n(\mathbf{r})  \times I_{\text{str}}(\mathbf{r},\alpha,\varphi)$ in real space corresponds to a mixing of the object's spatial frequencies $\mathbf{k}$ with the spatial frequency $\mathbf{k}_0$ in Fourier space. Hence, initially unobservable spatial frequencies $\mathbf{k} > \mathbf{k}_\text{max}$ are encoded in the microscope's OTF support.  Taking a set of linearly independent images and applying computational post-processing allows for the retrieval of this information. The required procedure will be outlined below for 3D-SIM. The resolution enhancement reads $(\mathbf{k}_0 + \mathbf{k}_\text{max}) / \mathbf{k}_\text{max}$ and reaches 2 for the diffraction limit $\mathbf{k}_0 = \mathbf{k}_\text{max}$.

In 3D SIM the $0^{th}$-order beam $e^{i k z}$ of the diffraction grating is added \cite{Gustafsson2008}. In addition to the lateral phase shifts $\pm\varphi_r$, we consider an \textit{axial} phase shift $\varphi_z$ on the central beam, introduced e.g. by an optical element placed on the optical axis. The electric field thus reads
\begin{equation}
E (x,y,z) = e^{i (k_x x + k_y y - \varphi_r) + i k_z z } +  e^{i (k z + \varphi_z)} + e^{-i (k_x x + k_y y - \varphi_r) +  i k_z z} \, ,
\label{eq:7}
\end{equation}
and the 3D intensity pattern $I_{\text{str}}(\mathbf{r}) = |E(x,y,z)|^2$ calculates to  
\begin{equation}
I_{\text{str}}(\mathbf{r}) = 3 + 2\cos[2 (k_x x + k_y y)+ 2 \varphi_r]  + 4\cos[(k_x x + k_y y) + \varphi_r] \cos[(k-k_z) z + \varphi_z] \, ,
\label{eq:8}
\end{equation}
This pattern contains seven spatial frequency components. 
Taking the Fourier transform $FT\{I(\mathbf{r})\} = I_0 \sum_{j=1}^{7}  e^{i  \varphi_j } \delta (\mathbf{k}-\mathbf{k}_j)$  yields seven delta peaks in Fourier space with phases $\varphi_j$ (which are combinations of $\varphi_r$ and $\varphi_z$). For $\pm 1$ diffraction orders propagating at the angles $\pm 60^{\circ}$ the positions read $\mathbf{k}_j = (k_\rho,k_z)_j = (0,0)k$, $(\sqrt{3}/2,1/2)k$, $(\sqrt{3}/2,-1/2)k$, $(-\sqrt{3}/2,1/2)k$, $(-\sqrt{3}/2,-1/2)k$, $(\sqrt{3},0)k$, $(-\sqrt{3},0)k$, with $k_\rho = (k_x^2 + k_y^2)^{1/2}$. These are shown in Fig.\,\ref{fig:3D-1}(c), where the axes are normalized to the wavenumber $k=2\pi/\lambda$. Using this 3D illumination, taking the Fourier transform of Eq.\,(\ref{eq:5}) and utilizing convolution theorems yields \cite{Gustafsson2008} 
\begin{equation}
FT  \{ \text{SIM}(\mathbf{r}) \} = H(\mathbf{k}) \times \sum_{j=1}^{7}  c_j \, e^{i  \varphi_j }  \,  \tilde{n}(\mathbf{k} - \mathbf{k}_j) \, ,
\label{eq:9}
\end{equation}
where the $c_j$ represent weights [see prefactors of cosines in Eq.\,(\ref{eq:8})]. Note that $\tilde{n}(\mathbf{k}) = FT\{ n(\mathbf{r}) \}$ contains the sought-after spatial frequency information of the unknown object under investigation. In a single image all seven components are superposed. To disentangle them at least seven independent 3D images (by measurement of z-stacks) are required. This is established by varying the phases $\varphi_r$ and $\varphi_z$ and creates the linear system $A \mathbf{n} = \mathbf{G}$, where the elements of the $7 \times 7$ matrix $A$ are given by the phase terms $e^{i  \varphi_j }$ of Eq.\,(\ref{eq:9}) and $\mathbf{n}$ denotes a vector with entries $\tilde{n}(\mathbf{k} - \mathbf{k}_j)$. The vector $\mathbf{G}$  possesses Eq.\,(\ref{eq:9}) as entries and the system is solved by  $ \mathbf{n} = A^{-1} \mathbf{G}$.

In practice a convenient choice of exactly seven independent images is not readily achieved \cite{Gustafsson2008}. As a result the components in Fig.\,\ref{fig:3D-1}(c) are first disentangled along the lateral direction by taking five linearly independent images with phases $\varphi_r = 0,\frac{2\pi}{5}, \frac{4\pi}{5} , \frac{6\pi}{5} , \frac{8\pi}{5}$ for a fixed $\varphi_z = 0$. The application of a $5 \times 5$ matrix $A_r^{-1} $ yields the intermediate components:
\begin{itemize}
	\item $\tilde{n}_1(\mathbf{k})= \tilde{n}(\mathbf{k} - [k,k])$\; , \;$\tilde{n}_4(\mathbf{k}) = \tilde{n}(\mathbf{k} - [-\sqrt{3}k,0])$\; , \;$\tilde{n}_5(\mathbf{k}) = \tilde{n}(\mathbf{k} - [\sqrt{3}k,0])$
	\item $\tilde{n}_2(\mathbf{k}) = \tilde{n}(\mathbf{k} - [-\frac{\sqrt{3}k}{2},-\frac{k}{2}]) + \tilde{n}(\mathbf{k} - [\frac{\sqrt{3}k}{2},-\frac{k}{2}])$
	%\item $n_2(\mathbf{k}) = n(\mathbf{k} - [-\sqrt{3}k/2,-k/2]) + n(\mathbf{k} - [\sqrt{3}/2,-k/2])$
	\item $\tilde{n}_3(\mathbf{k}) = \tilde{n}(\mathbf{k} - [-\frac{\sqrt{3}k}{2},\frac{k}{2}]) + \tilde{n}(\mathbf{k} - [\frac{\sqrt{3}k}{2},\frac{k}{2}])$\,,
	%\item $n_3(\mathbf{k}) = n(\mathbf{k} - [-\sqrt{3}k/2,k/2]) + n(\mathbf{k} - [\sqrt{3}/2,k/2)]$
	%\item $\tilde{n}_4(\mathbf{k}) = \tilde{n}(\mathbf{k} - [-\sqrt{3}k,0])$
	%\item $n_4(\mathbf{k}) = n(\mathbf{k} - [-\sqrt{3}k,0])$
	%\item $\tilde{n}_5(\mathbf{k}) = \tilde{n}(\mathbf{k} - [\sqrt{3}k,0])$ \ ,
	%\item $n_5(\mathbf{k}) = n(\mathbf{k} - [\sqrt{3}k,0])$
\end{itemize}
where $\tilde{n}_1(\mathbf{k})$, $\tilde{n}_4(\mathbf{k})$ and $\tilde{n}_5(\mathbf{k})$ are already fully isolated, and $\tilde{n}_2(\mathbf{k})$ and $\tilde{n}_3(\mathbf{k})$ contain two components each. To disentangle them an additional measurement series (with 5 different $\varphi_r$) for $\varphi_z = \frac{2\pi}{3}$, and application of a $2 \times 2$ matrix $A_z^{-1} $, is required. In summary the flow chart requires $2 \times 5 = 10$ images. Note that, alternatively to varying the axial phase $\varphi_z $ one can regard the axial modulation to act on the PSF and keep it fixed with respect to the objective coordinates axes \cite{Gustafsson2008}. Correspondingly the axial OTF support is enhanced two-fold and requires a two-fold finer sampling of focal planes to maintain the Nyquist sampling rate. The required amount of images is hence the same for both approaches.

To sufficiently cover the enlarged OTF support the procedure needs to be repeated for three orientations $\alpha = \tan^{-1}(k_y/k_x) = 0,\frac{1\pi}{3}, \frac{2\pi}{3}$ [see the final image in Fig.\,\ref{fig:3D-1}(d)]. In the next step the disentangled components need to be shifted back to their true positions in Fourier space, post-processed appropriately and merged into a large homogenous support. This is achieved trough application of the formula \cite{Gustafsson2008}
\begin{equation}
\tilde{n}_\text{new}(\mathbf{k}) = \frac{\sum_j \tilde{n}_\text{j}(\mathbf{k}+\mathbf{k}_j)}{\left[\sum_j H(\mathbf{k}+\mathbf{k}_j) \right] + \gamma} A(\mathbf{k}) \, ,
\label{eq:SIM1}
\end{equation}
where $\gamma$ is a constant that prevents division by zero and should be chosen noise-dependently. $A(\mathbf{k})$ is a triangular apodization function in 3D and serves the purpose of reducing ringing in the final image, which is obtained via inverse Fourier transformation of Eq.\,(\ref{eq:SIM1}). The deconvolution in Eq.\,(\ref{eq:SIM1}) also enhances the suppressed spatial frequency information from the outer rims of the OTF support and thus leads to the highest possible resolution for the given data. The straightforward assembly in Fourier space requires near-integer numbers $\mathbf{k}_j = (k_x,k_y,k_z)_j$ \cite{Gustafsson2000,Gustafsson2008} such that the assembly is typically conducted in real space by first applying the inverse Fourier transform to each component and then multiplying by the complex wave $e^{i \mathbf{k}_j \mathbf{r}}$. Note that while in 3D-SIM the resolution is doubled along all axes, it is  still limited by diffraction. 

\begin{figure}[h]%eq:2a
\centering
\includegraphics[width=1.0 \linewidth]{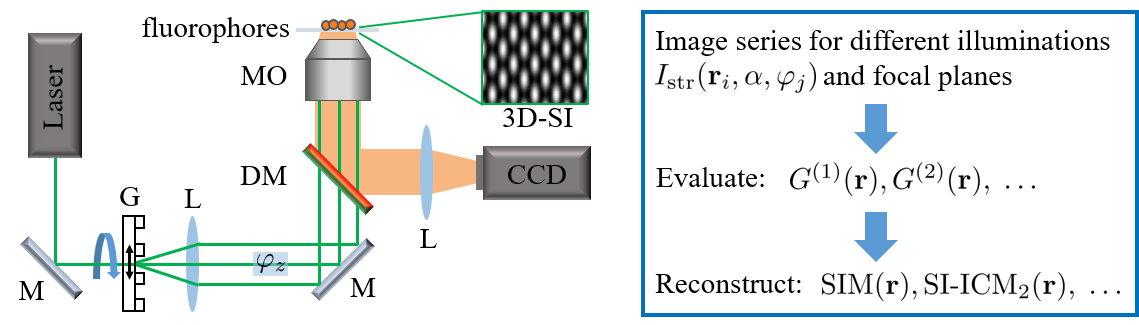}
\caption{Schematic setup of an SI-ICM experiment (left side) and the corresponding flowchart to obtain the sought-after superresolving images (right side). For details see text.}
\label{fig:setup}%
\end{figure}

SI-ICM fruitfully combines SIM and ICM. That is, the structured illumination encodes information from outside the original OTF support via spatial frequency mixing and and the correlation analysis effectively raises all signals to the $m$th power. The schematic setup and the experimental flowchart are shown in Fig.\,\ref{fig:setup}. A laser illuminates a diffraction grating which produces the diffraction orders $-1,0,+1$. After collimation by the lens L the three beams are coupled into the back focal of a microscope objective (MO) and form the three dimensional structured illumination (3D-SI). Rotation and translation of the grating varies the orientation $\alpha$ and the lateral phase $\varphi_r$, respectively. An additional optical element on the optical axis varies the axial phase $\varphi_z$. The fluorescence emission from the fluorophores is captured by the same MO and guided toward an CCD camera, which captures an image series for each $ I_{\text{str}}(\mathbf{r}_i,\alpha,\varphi_j)$. The basic flow chart in the experiment would be: i) set a specific value set $(\varphi_r,\varphi_z,\alpha)$ for the 3D-SI, ii) take a 2D image series and evaluate it according to ICM algorithms [cf. Eq.\,(\ref{eq:2a})], iii) repeat the second step for varying $\varphi_r$, iv) repeat steps two and three for different focal planes to obtain a z-stack, also with a sufficient number of values $\varphi_z$ per focal plane, v) repeat steps two to four for the next pattern orientation $\alpha$, and vi) apply a SIM reconstruction algorithm to the set of 3D ICM images with different illumination pattern values $(\varphi_r,\varphi_z,\alpha)$.

In mathematical terms the outlined procedure corresponds to a combination of Eqs.\,(\ref{eq:3}) and (\ref{eq:5}) which results in
\begin{equation}
\text{SI-ICM}_m(\mathbf{r}) =  \sum_{i=1}^N  h_m(\mathbf{r}-\mathbf{r}_i)  \times I_{\text{str}}(\mathbf{r}_i,\alpha,\varphi_j)^m  \, .
\label{eq:6}
\end{equation}
Now, higher harmonics up to $\cos (m \mathbf{k}_0 \mathbf{r})$ arise and the individual OTF $H_m(\mathbf{k})$ is enlarged by the factor $m$ (when including deconvolution). For 2D SIM  combined with a correlation analysis the lateral resolution was shown to be enhanced by up to $m + m = 2m$ \cite{Classen2017a,Zhao2017}. 

To illustrate the outcome of Eq.\,(\ref{eq:6}) we consider the 3D structured illumination of Eq.\,(\ref{eq:8}) and second-order SI-ICM ($m=2$). Taking the Fourier transform of Eq.\,(\ref{eq:6}) thus yields
\begin{equation}
FT  \{ \text{SI-ICM}_2(\mathbf{r}) \} = H_2(\mathbf{k}) \times \sum_{j=1}^{19}  c_j \, e^{i  \varphi_j }  \,  \tilde{n}(\mathbf{k} - \mathbf{k}_j) \, ,
\label{eq:10}
\end{equation}
where 19 spatial frequency components arise, which can readily be calculated by executing $I_\text{str}(\mathbf{r})^2= (|E(x,y,z)|^2)^2$. Higher harmonics now arise along the lateral and axial direction, as illustrated in Fig.\,\ref{fig:3D-1}(f) by the added most outer (red) dots. Moreover, the OTF is enlarged by the factor $2$ [see Fig.\,\ref{fig:3D-1}(e)]. The total improvement reaches up to $2 + 2 = 4$ along all axes [see the 3D volume in Fig.\,\ref{fig:3D-1}(g)]. The larger OTF $H_2(\mathbf{k})$ also reduces the need for many orientations $\alpha$ as compared to saturated SIM \cite{Gustafsson2005} so four $\alpha = 0,\frac{1\pi}{4}, \frac{2\pi}{4} , \frac{3\pi}{4}$ are sufficient \cite{Classen2017a}.

In principle 19 independent 3D images would suffice, yet a convenient choice is not readily achieved. Hence, as above, in the first step nine images with different lateral phases $\varphi_r = \frac{2\pi}{9} j $ ($j = 0, \ldots, 8$) need to be acquired to disentangle the components along the lateral direction. By this, only the lateral components with center positions $(2\sqrt{3},0)k$, $(-2\sqrt{3},0)k$ [see Fig.\,\ref{fig:3D-1}(f)] are isolated. The remaining contributions are still composed of two or three individual axial components. Hence the first step needs to be repeated for three axial phases $\varphi_z =  0,\frac{2\pi}{3}, \frac{4\pi}{3}$, resulting in $3 \times 9 = 27$ measurements per orientation $\alpha$. Including the four orientations $\alpha$ a total of 108 images is obtained.

\section{Simulation}
\label{sec:simu}

\begin{figure}[b]%
\centering
\includegraphics[width=0.9 \linewidth]{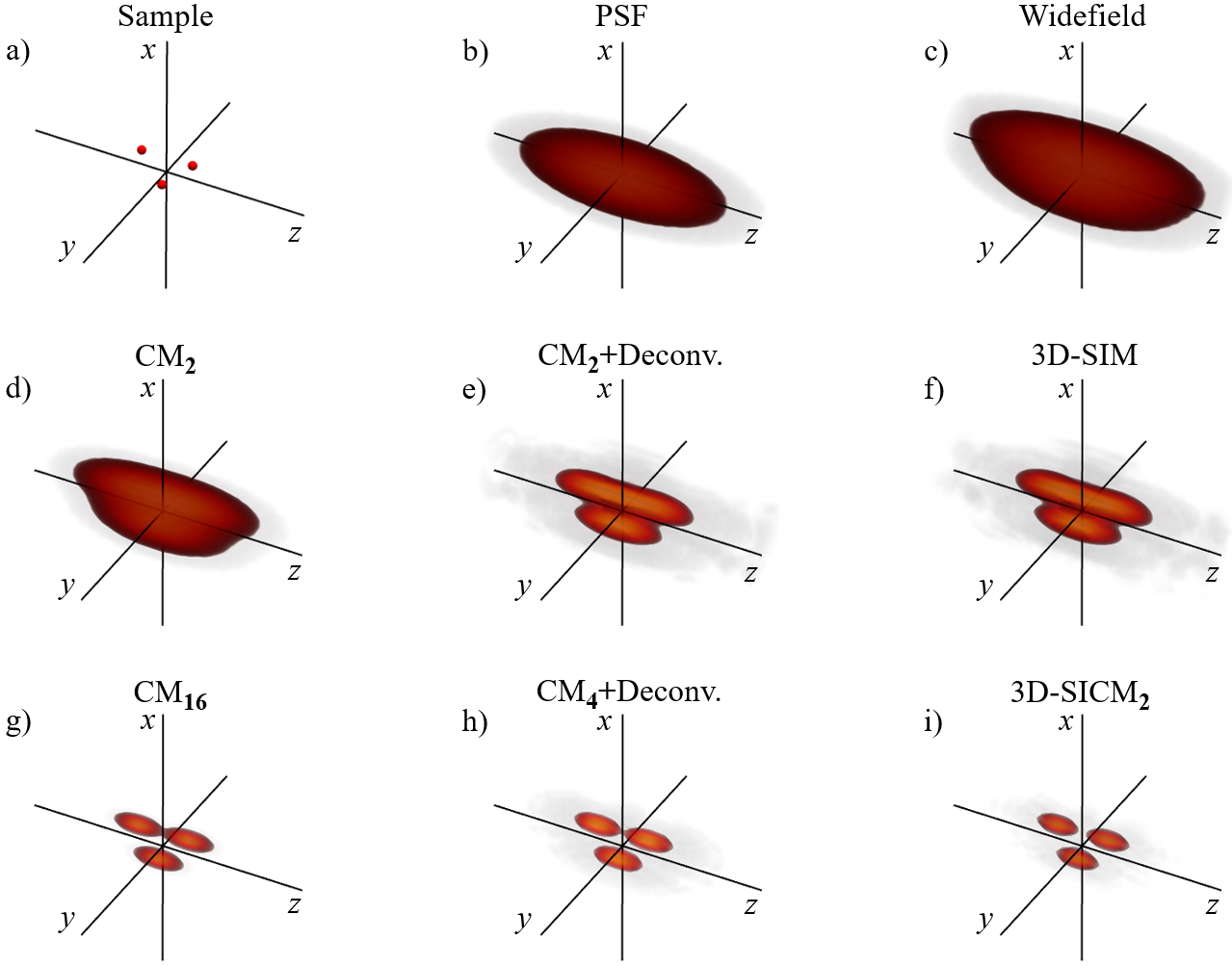}
\caption{The figure shows (a) an object  consisting of three emitters at positions $\mathbf{r}_1=(-0.16,0.16,0.05)$, $\mathbf{r}_2=(0.26,-0.26,0.57)$ and $\mathbf{r}_3=(0.26,-0.26,0.68)$ (in units of $\Delta \rho_\text{min}$) and (b) the 3D PSF of Eq.\,(\ref{eq:1}) utilized in the simulation. The images (c) - (i) are obtained by the methods (c) widefield microscopy, (d) second-order ICM, (e) second-order ICM + Deconvolution, (f) 3D-SIM, (g) $16^\text{th}$-order ICM, (h) fourth-order ICM + Deconvolution, and (i) second-order 3D-SI-ICM. For details on the simulation see text.}
\label{fig:3D-3}%
\end{figure}

To illustrate the mathematical description of section 2, we performed a basic simulation of the formulas  in Eqs.\,(\ref{eq:2b}), (\ref{eq:3}), (\ref{eq:5}) and (\ref{eq:6}), where we utilized the 3D PSF of Eq.\,(\ref{eq:1}) [see Fig.\,\ref{fig:3D-3}(b)]. In case of Eq. (\ref{eq:3}) we consider the direct result with a resolution enhancement of $\sqrt{m}$ and after deconvolution via a Wiener filter and triangular apodization, resulting in an enhancement of up to factor $m$ [see e.g. Fig.\,\ref{fig:3D-3}(d) and \ref{fig:3D-3}(e) for second-order correlation and Fig.\,\ref{fig:3D-3}(h) for fourth-order correlation]. Since we are interested in the resolution power of ICM, SIM and SI-ICM relative to widefield microscopy we use dimensionless units and merely normalize $\mathbf{r}$ by the Rayleigh limit $\Delta \rho_\text{min}$ of Eq.\,(\ref{eq:2}).

We chose a setup with three close-by emitters distributed in the 3D object space within a sub-diffraction limited volume [see Fig.\,\ref{fig:3D-3}(a)]. The coordinates $\mathbf{r}_1=(-0.16,0.16,0.05)$, $\mathbf{r}_2=(0.26,-0.26,0.57)$ and $\mathbf{r}_3=(0.26,-0.26,-0.68)$ with pair-wise lateral or axial separations were chosen to demonstrate the full 3D resolution capabilities of SI-ICM. For illustration purposes we used theoretical data without noise. A discussion of possible practical limitations and requirements is given below. The 3D data was calculated with respect to the 3D PSF of Eq.\,(\ref{eq:1}). In the experiment it would be obtained through the flow chart outlined in section 2. For the deconvolution post-processing we always chose $\gamma = 0.001$ for the methods ICM, SIM and SI-ICM. As outlined above, for SIM five lateral phases $\varphi_r$, two axial phases $\varphi_z$ and three orientations $\alpha$ are required, resulting in a total of 30 images.For $\text{SI-ICM}_2$ the required phases are nine $\varphi_r$, three $\varphi_z$ and for four $\alpha$, yielding a total of 108 images. 

Fig.\,\ref{fig:3D-3} shows the simulations results for seven different signals. These are (c) widefield microscopy, (d) second-order ICM, (e) second-order ICM + Deconvolution, (f) 3D-SIM, (g) $16^\text{th}$-order ICM, (h) fourth-order ICM + Deconvolution, and (i) second-order 3D-SI-ICM. The image in (g) was chosen for comparison purposes, since it provides a four-fold resolution enhancement over regular widefield microscopy due to the PSF being taken to the 16th power. The simulation results are in good agreement with the theory and it can be seen that 3D SI-ICM equally enhances the lateral and axial resolution and thus provides full 3D superresolution. Further it can be seen that second-order SI-ICM achieves the same resolution enhancement as fourth-order ICM + Deconvolution, what can be a major advantage since the evaluation of high correlation orders is challenging. 

For widefield SI-ICM in 2D and 3D the question is whether the approach will show superior performance in experiments compared to the already existing techniques. The final answer can of course only be given by future experiments, but an estimation of the expected requirements might guide the way. Toward this we compare the required amount of frames for $\text{SI-ICM}_2$ and $\text{ICM}_4$, for the SOFI variant with blinking fluorophores. First, for 3D $\text{SI-ICM}_2$ we showed that 108 images series are required per focal plane. For $\text{ICM}_4$ a single image series (however with more frames) is sufficient for a single transverse plane. To account for the increased axial resolution the sampling of focal planes is required to be four times as dense to fulfill the Nyquist rate. For imaging a 3D volume the difference in image series results in the factor 27.

Yet, the required amount of frames per series differs significantly with rising correlation order \cite{Geissbuehler2011,Stein2017}. Further it strongly depends on the imaging scenario, that is the blinking dynamics and the respective on-time to off-time ratio. The emitter density within a diffraction limited volume also strongly affects the convergence of the (higher-order) cumulants toward their theoretical values \cite{Stein2017}. Generally the second-order cumulant converges within 100-500 frames (depending on the specific scenario). The fourth-order cumulant requires around $\sim 15$ times more frames in the best case to converge, but this value quickly diverges, for example, when the emitter density increases. The second-order cumulant, by contrast, quickly approaches an asymptotic limit and can be used with very high density samples \cite{Stein2017}. An additional major advantage is the smaller brightness skewing of the second-order cumulant compared to higher orders.

Once a 3D ICM image stack with different illumination patterns $I_{\text{str}}(\mathbf{r}_i,\alpha,\varphi_j)$ is obtained with a certain signal to noise ratio a SIM reconstruction algorithm needs to reconstruct the final 3D SI-ICM image. Hereby the question arises how the performance and resolution of this algorithm scales with a given signal to noise ratio. Since this question has been answered elsewhere, we refer the reader to \cite{Classen2017a,Zhao2017} or other more detailed studies \cite{Ingerman2018}.

Finally, we point out that a first proof-of-principle experiment that relies on the SI-ICM principle, while in a confocal microscopy setting, was recently demonstrated by Tenne \textit{et al.} \cite{Tenne2018}. In the paper the authors combine \textit{image scanning microscopy} (which can be regarded as a confocal SIM variant \cite{Strohl2016}) with the evaluation of quantum correlations by making use of antibunching of individual quantum dots. While their setting is different from the widefield microscopy setup discussed here it delivers a first cornerstone towards real applications. 

Regarding the generally high amount of frames, we point out that very fast cameras exist, which possess frame rates as high as a few kHz. Even modern EM-CCD cameras such as the Andor iXon 897 achieve these frame rates using the cropped mode. Considering a total amount of frames $200 \times 108 \times 10 \approx 2 \times 10^5$, the measurement time would equal a few tens of seconds for a full 3D superresolution image with 10 focal planes. With improving detector technology this value can be foreseen to be  significantly smaller in the future.

\section{Conclusion}

In this paper we proposed to enhance the resolution power of ICM through the addition of 3D structured illumination. We presented a mathematical treatment that predicts a resolution enhancement of $m+m = 2m$  through SI-ICM with correlation order $m$, both along the lateral and axial direction. Compared to the enhancement factor $m$ of ICM alone this is a major boost that allows to reach deep-subwavelength resolution already with much lower correlation orders. Further, we outlined the flow chart for an experiment and illustrated the results via basic simulations that matched the theoretical predictions. Moreover we pointed out that SIM in combination with second-order ICM requires an comparable amount of images as fourth-order ICM itself and and would outperforms each method on its own. We note that since SI-ICM fully operates within the linear regime it bears the potential to increase the resolution in particular for imaging biological specimen at low illumination levels, i.e., especially in cases where other methods can not be utilized.

Further enhancements of the axial resolution can be achieved by combining SI-ICM with the double-objective 3D-SIM technique known as $\text{I}^5$S \cite{Shao2008}. The three added coherent beams from the second objective lead to very fast modulations along the axial direction. Additionally, the OTF is enlarged along the axial direction as in 4Pi- and $\text{I}^5$-microscopy \cite{Hell1994a,Gustafsson1999}. Again, SI-ICM would square the PSF and the excitation pattern. 
 
Another promising future route may be to combine SI-ICM with plasmonic SIM techniques \cite{Zubairy2014,Wei2014,Ponsetto2017}. Even though these techniques are limited to 2D, they allow for spatial frequencies $\mathbf{k}_{0} > \mathbf{k}_\text{max}$ of the standing wave pattern. In linear plasmonic SIM $\mathbf{k}_{0} = 2\mathbf{k}_\text{max}$ should not be exceeded to prevent gaps in the OTF support coverage. The resolution enhancement is thus limited to the factor 3 \cite{Wei2014,Ponsetto2017}. SI-ICM, however, would highly benefit from spatial frequencies $\mathbf{k}_0 > 2 \mathbf{k}_\text{max}$ since the enlarged OTF $H_m(\mathbf{k})$ prevents an early formation of gaps and the higher harmonics $\cos (m \mathbf{k}_0 \mathbf{r})$ reach out to very high spatial frequencies. 

\section*{Funding Information}

A.C. and J.v.Z: Erlangen Graduate School in Advanced Optical Technologies (SAOT)
; G.S.A.: Welch Foundation (Award number: A-1943-20180324).

\section*{Acknowledgments}

A.C. gratefully acknowledges the hospitality of Texas A\&M University where parts of this work were done.

\bibliography{Refs_SI-ICM}

\end{document}